\documentclass[%
 preprint,
 amsmath,amssymb,
 aps,
prb,
]{revtex4-1}

\usepackage{graphicx}
\usepackage{mwe}
\usepackage{float}
\usepackage{bm}
\begin{document}

\title{Crystal-symmetry-based selection rules for anharmonic phonon-phonon scattering from a group theory formalism}

\author{Runqing Yang}
\affiliation{Department of Mechanical Engineering, University of California, Santa Barbara, CA 93106, USA} 

\author{Shengying Yue}
\affiliation{Department of Mechanical Engineering, University of California, Santa Barbara, CA 93106, USA}

\author{Yujie Quan}
\affiliation{Department of Mechanical Engineering, University of California, Santa Barbara, CA 93106, USA}

\author{Bolin Liao}
\email{bliao@ucsb.edu}
\affiliation{Department of Mechanical Engineering, University of California, Santa Barbara, CA 93106, USA}


\date{\today}

\begin{abstract}

Anharmonic phonon-phonon scattering serves a critical role in heat conduction in solids. Previous studies have identified many selection rules for possible phonon-phonon scattering channels imposed by phonon energy and momentum conservation conditions and crystal symmetry. However, the crystal-symmetry-based selection rules have mostly been \textit{ad hoc} so far in selected materials, and a general formalism that can summarize known selection rules and lead to new ones in any given crystal is still lacking. In this work, we apply a general formalism for symmetry-based scattering selection rules based on the group theory to anharmonic phonon-phonon scatterings, which can reproduce known selection rules and guide the discovery of new selection rules between phonon branches imposed by the crystal symmetry. We apply this formalism to analyze the phonon-phonon scattering selection rules imposed by the in-plane symmetry of graphene, and demonstrate the significant impact of symmetry-breaking strain on the lattice thermal conductivity. Our work quantifies the critical influence of the crystal symmetry on the lattice thermal conductivity in solids and suggests routes to engineer heat conduction by tuning the crystal symmetry. 

\end{abstract}

\maketitle


\section{Introduction}
Phonon scattering processes play a central role in the atomic-level understanding of thermal transport in semiconductors and insulators\cite{lindsay2016first}. The intrinsic anharmonic phonon-phonon scattering, which includes three-phonon, four-phonon and higher-order phonon scattering processes\cite{feng2020higher}, is the mechanism that dominates the behavior of the lattice thermal
conductivity for crystalline semiconductors and insulators around and above room temperature. Extensive theoretical, computational\cite{lindsay2016first} and experimental\cite{hua2020phonon} efforts have focused on understanding the anharmonic phonon-phonon scattering processes in detail. Maradudin et al.\cite{maradudin1962scattering,maradudin1962on} established the anharmonic lattice
dynamics framework to predict intrinsic three-phonon
scattering rates in solids\cite{broido2007intrinsic} by creating the third-order anharmonic Hamiltonian and adopting the phonon creation and annihilation operators in the second quantization formalism of many-body physics. Recent advancements of first-principles methods, such as the density functional perturbation theory (DFPT)\cite{baroni2001phonons} and the finite-displacement approach\cite{broido2007intrinsic} have enabled routine calculations of the anharmonic three-phonon scattering rates in realistic single crystalline materials. Feng and Ruan\cite{feng2016quantum} further generalized the first-principles calculation to include the four-phonon scattering process. Combining first-principles calculations of phonon scattering rates with the phonon Boltzmann transport equation (BTE) solvers, the lattice thermal conductivity of a wide range of crystalline materials can now be computed with a high accuracy, often showing excellent agreements with experiments\cite{turney2009predicting,lindsay2009lattice,lindsay2012thermal,yue1,yue2}.

In addition to the advancement of first-principles computational methods, an improved physical understanding of the phonon-phonon scattering, particularly the selection rules that determine allowed and forbidden scattering channels in a given material, has led to the rational discovery of materials with desirable thermal transport properties. Many of these selection rules are imposed by the requirement of energy and momentum conservation during phonon-phonon scattering events\cite{ravichandran2020phonon}. One well-known example is the large acoustic-optical band gaps typically existing in materials with a large mass contrast between constituent atoms that forbid the $aao$ type of phonon scattering processes involving two acoustic phonons and one optical phonon\cite{lindsay2013first}. Another example is the ``acoustic-bunching'' effect, where overlapping acoustic branches in the phonon dispersion limit the $aaa$ scattering channels involving three acoustic phonons\cite{Ziman1960}. Both selection rules have contributed to the unexpected high lattice thermal conductivity in boron arsenide\cite{lindsay2013first} that has been experimentally verified recently\cite{tian2018unusual,li2018high,kang2018experimental}.  

Besides the energy and momentum conservation conditions, the crystal symmetry also places restrictions on possible phonon scattering channels. For example, symmetry-based analysis of the selection rules have been extensively used to interpret optical spectroscopic measurements involving photon-phonon interactions, such as the Raman and infrared spectroscopy\cite{dresselhaus2007group}. The impact of symmetry-based selection rules on electron-phonon scatterings has also been examined to understand the strain-engineering of charge mobility in semiconductors\cite{baykan2010strain,sun2007physics} and the electrical transport in two-dimensional (2D) materials\cite{castro2010limits}. In the context of phonon properties and thermal transport, crystal symmetry has been used to reduce the number of unique interatomic force constants (IFCs) in first-principles phonon calculations and improve their numerical accuracy\cite{esfarjani2008method,togo2015first}. In addition, crystal symmetry can place contraints on the phonon dispersion relations, which indirectly affect phonon scattering through energy and momentum conservation conditions\cite{phonongra}. Furthermore, phonon-phonon scattering rules that are directly imposed by crystal symmetry have been studied in a few material systems with special crystal symmetries. For example, the existence of a screw axis in certain one-dimensional (1D) systems, such as the carbon nanotubes\cite{lindsay2009lattice} and the 1D-chain compound Ba$_3$N,\cite{pandey2018symmetry} imposes an additional phonon scattering selection rule requiring the conservation of a phonon angular momentum. As another example, Lindsay et al.\cite{lindsay2010flexural} discussed the selection rule for three-phonon scatterings involving flexural phonon modes associated with the out-of-plane mirror reflection symmetry in 2D materials, which contributes to the high lattice thermal conductivity of graphene. However, most crystalline materials possess more crystal symmetries and should have more phonon scattering selection rules imposed by their full crystal symmetry properties. A natural development, therefore, is to systematically identify phonon scattering selection rules for a given crystal based on its full crystal symmetry.

Group theory provides a systematic mathematical framework to analyze the impact of crystal symmetries on phonon-phonon scattering. To reveal additional phonon scattering selection rules, we first review the three-phonon scattering matrix elements written in complex normal coordinates for phonon eigenmodes  \cite{Bornhuang,latticedynamics}. We then investigate the symmetry properties of the complex normal coordinates \cite{latticedynamics} and apply the method of space group selection rules \citep{MEB,BCC} to derive the phonon scattering selection rules in crystals with symmorphic space groups. We provide several examples to illustrate how those selection rules can be identified based on the group theory and provide justifications for our findings. Our results not only show good compatibility with the findings in previous research\cite{lindsay2010flexural,pandey2018symmetry}, but also lead to additional selection rules for phonon scattering channels that have not been discussed before. To demonstrate the effect of these selection rules, we compare the phonon scattering rates of forbidden channels and allowed channels in graphene with or without a symmetry-breaking strain using first-principles simulations. The simulation results agree well with our theoretical predictions and show that the group theory framework we have generalized can serve as a valuable tool to search for phonon-phonon scattering selection rules in crystalline materials and investigate the impact of crystal symmetry on phonon transport. Our result also implies a potential route towards tuning the thermal conductivity of solids via controlled crystal-symmetry breaking. We note that the group-theory formalism has been applied to discuss the thermal conductivity of transition metal dichalcogenides by Cammarata\cite{cammarata2019phonon}, however, the scattering selection rules for phonons with particular wavevectors were not established. 

\section{three-phonon scattering rates in normal coordinates}
The Hamiltonian of a crystal can be written as the sum of the harmonic part $H_0$ and the anharmonic parts\cite{Bornhuang,scatterderi,feng2016quantum}:
\begin{equation} \label{eqn:1}
\textit{H}=\textit{H}_0+\textit{H}_3+\textit{H}_4+\dots   
\end{equation}
The third-order terms are expressed as: 
\begin{equation} \label{eqn:2}
\begin{split}
 \textit{H}_3=\frac{1}{6}\sum_{l,\kappa,\alpha}   \sum_{l',\kappa',\beta}\sum_{l'',\kappa'',\gamma}\Phi_{\alpha \beta \gamma}(l\kappa, l{'}\kappa{'}, l{''}\kappa{''})
 u_{\alpha}\binom{l}{\kappa}u_{\beta}\binom{l'}{\kappa'}u_{\gamma}\binom{l''}{\kappa''}.
\end{split}
\end{equation}
Here $\Phi$ represents the force constant and $u$ is the atomic displacement. $\alpha$, $\beta$ and $\gamma$ are Cartesian coordinates. $l$,$l'$,$l''$ and $\kappa$,$\kappa'$,$\kappa''$  represent the indices for a certain unit cell and a particular atom within that cell. The atomic displacements $u$ can be expanded in the basis of the phonon eigenvectors \cite{Bornhuang,latticedynamics}: 

\begin{equation} \label{eqn:1}
\begin{split}
    u_{\alpha}\binom{l}{\kappa}=\frac{1}{\sqrt{N}\sqrt{M_\kappa}} \sum_{\textit{\textbf{q}}}\sum_{j,{\rho}}^{} Q \binom{\textit{\textbf{q}}}{j_\rho} e_{\alpha}^{\kappa} \binom{\textit{\textbf{q}}}{j_\rho}e^{i \textit{\textbf{q}} \cdot \textit{\textbf{R}}_{l}}.
\end{split}
\end{equation} 
Here $N$ is the number of atoms, $M_{\kappa}$ is the mass of the atom $\kappa$. $\mathbf{q}$ is the phonon wavevector and $j$ is the phonon branch index. $\rho$ is an extra index labeling degenerate phonon modes for a given branch. $\rho =1,\dots, l_j$ and $l_j$ is the number of degenerate eigenvectors. $e_{\alpha}^{\kappa} \binom{\textit{\textbf{q}}}{j_\rho}$ is one component of the eigenvector of the phonon mode $(\mathbf{q},j_\rho)$ associated with the displacement of atom $\kappa$ along the direction $\alpha$. $\mathit{\mathbf{R}_{l}}$ is the coordinate of a unit cell. The expansion coefficients $Q \binom{\textit{\textbf{q}}}{j_\rho}$ are the complex normal coordinates for the phonon eigenmodes.
We can transform Eq. (\ref{eqn:2}) into the following form by rewriting the atomic displacements $u_{\alpha}\binom{l}{\kappa}$ in normal coordinates $Q \binom{\textit{\textbf{q}}}{j_\rho}$ 
\begin{equation} \label{eqn:4}
\begin{split}
 \textit{H}_3=&\frac{1}{6N^{\frac{3}{2}}}\sum_{\textit{\textbf{q}},j,{\rho}}\sum_{\textit{\textbf{q}}',j',{\rho'}}\sum_{\textit{\textbf{q}}'',j'',{\rho''}}Q \binom{\textit{\textbf{q}}}{j_\rho}Q \binom{\textit{\textbf{q}}'}{j'_{\rho'}}Q \binom{\textit{\textbf{q}}''}{j''_{\rho''}}
 \\\times &\sum_{l,\kappa,\alpha}   \sum_{l',\kappa',\beta}\sum_{l'',\kappa'',\gamma} \frac{\Phi_{\alpha \beta \gamma}(l\kappa, l{'}\kappa{'}, l{''}\kappa{''})}{\sqrt{M_\kappa M_{\kappa'} M_{\kappa''}}}\\
 \times&e_{\alpha}^{\kappa} \binom{\textit{\textbf{q}}}{j_{\rho}}e_{\beta}^{\kappa'} \binom{\textit{\textbf{q}}'}{j'_{\rho'}}e_{\gamma}^{\kappa''} \binom{\textit{\textbf{q}}''}{j''_{\rho''}} e^{i \textit{\textbf{q}} \cdot \textit{\textbf{R}}_{l}+i \textit{\textbf{q}}' \cdot \textit{\textbf{R}}_{l'}+i \textit{\textbf{q}}'' \cdot \textit{\textbf{R}}_{l''}}.
\end{split}
\end{equation}
Eq.(\ref{eqn:4}) can be rewritten in a more compact form for the phonon absorption case (two phonons merge into a third phonon):
\begin{align}\label{eqn:5}
\begin{split}
  \textit{H}_3=\frac{1}{6N^{\frac{1}{2}}}\sum_{\textit{\textbf{q}},j,{\rho}}\sum_{\textit{\textbf{q}}',j',{\rho'}}\sum_{\textit{\textbf{q}}'',j'',{\rho''}}&V_+(\textit{\textbf{q}}j_{\rho},\textit{\textbf{q}}'j'_{\rho'}, \textit{\textbf{q}}''j''_{\rho''})\Delta(\textit{\textbf{q}}+ \textit{\textbf{q}}'- \textit{\textbf{q}}'')\\
  \times &Q \binom{\textit{\textbf{q}}}{j_{\rho}}Q \binom{\textit{\textbf{q}}'}{j'_{\rho'}}Q \binom{\textit{\textbf{q}}''}{j''_{\rho''}}^* ,
\end{split}
\end{align}
where $V_+(\textit{\textbf{q}}j_{\rho},\textit{\textbf{q}}'j'_{\rho'}, \textit{\textbf{q}}''j''_{\rho''})$ and $\Delta(\textit{\textbf{q}}+ \textit{\textbf{q}}'- \textit{\textbf{q}}'')$ are: 
\begin{align}\label{eqn:6}
\begin{split}
   V_+(\textit{\textbf{q}}j_{\rho},\textit{\textbf{q}}'j'_{\rho'}, \textit{\textbf{q}}''j''_{\rho''})&= 
 \sum_{\kappa,\alpha}   \sum_{l,\kappa',\beta}\sum_{l',\kappa'',\gamma} \frac{\Phi_{\alpha \beta \gamma}(0\kappa, l\kappa{'}, l{'}\kappa{''})}{\sqrt{M_\kappa M_{\kappa'} M_{\kappa''}}}\\
\times&e_{\alpha}^{\kappa} \binom{\textit{\textbf{q}}}{j_{\rho}}e_{\beta}^{\kappa'} \binom{\textit{\textbf{q}}'}{j'_{\rho'}}e_{\gamma}^{\kappa''} \binom{\textit{\textbf{q}}''}{j''_{\rho''}}^* e^{i \textit{\textbf{q}}' \cdot \textit{\textbf{R}}_{l}-i \textit{\textbf{q}}''\cdot \textit{\textbf{R}}_{l'}}
\end{split}
\end{align}
\begin{align}\label{eqn:7}
\Delta(\textit{\textbf{q}}+ \textit{\textbf{q}}'- \textit{\textbf{q}}'')=\left\{
\begin{array}{rcl}
1       &      & {\textit{\textbf{q}}+ \textit{\textbf{q}}'- \textit{\textbf{q}}''=\textit{\textbf{K}}}\\
0     &      & {\textrm{otherwise}}
\end{array} \right.
\end{align}
where $\textit{\textbf{K}}$ can be any reciprocal lattice vector.
Eq.(\ref{eqn:6}) and Eq.(\ref{eqn:7}) are extensively used in first-principles phonon scattering calculations\cite{scatterderi}. The three-phonon absorption rate $\Gamma_{\textit{\textbf{q}}j_{\rho},\textit{\textbf{q}}'j'_{\rho'}, \textit{\textbf{q}}''j''_{\rho''}}^{+}$ can be expressed as \cite{shengbte01}:
\begin{equation}\label{eqn:8}
\begin{split}
\Gamma_{\textit{\textbf{q}}j_{\rho},\textit{\textbf{q}}'j'_{\rho'}, \textit{\textbf{q}}''j''_{\rho''}}^{+}=\frac{\hbar \pi}{4} \frac{f_{0}^{\prime}-f_{0}^{\prime \prime}}{\omega_{\textit{\textbf{q}}j_{\rho}} \omega_{\textit{\textbf{q}}'j'_{\rho'}} \omega_{\textit{\textbf{q}}''j''_{\rho''}}} |V_+(\textit{\textbf{q}}j_{\rho},\textit{\textbf{q}}'j'_{\rho'}, \textit{\textbf{q}}''j''_{\rho''})|^2
\delta\left(\omega_{\textit{\textbf{q}}j_{\rho}}+\omega_{\textit{\textbf{q}}'j'_{\rho'}}-\omega_{\textit{\textbf{q}}''j''_{\rho''}}\right)
\end{split}
\end{equation}
where $f_0$
stands for $f_0(\omega_\lambda)$, the Bose–Einstein distribution, and $V_+(\textit{\textbf{q}}j_{\rho},\textit{\textbf{q}}'j'_{\rho'}, \textit{\textbf{q}}''j''_{\rho''})$ can be viewed as the absorption scattering matrix element of three specific phonons. The two delta functions in Eq.(\ref{eqn:7}) and Eq.(\ref{eqn:8}) impose the conservation of energy and momentum during the scattering process. Eq.(\ref{eqn:5}) shows the third-order anharmonic Hamiltonian can be expressed as a linear combination of triple products of the complex normal coordinates $Q$ with the coefficient $V_+(\textit{\textbf{q}}j_{\rho},\textit{\textbf{q}}'j'_{\rho'}, \textit{\textbf{q}}''j''_{\rho''}) \Delta(\textit{\textbf{q}}+ \textit{\textbf{q}}'- \textit{\textbf{q}}'')$.
In order to derive the scattering selection rules in the framework of the group theory, we need to find the transformation properties of the left-hand side and the right-hand side of Eq.(\ref{eqn:5}) from the symmetry perspective. First we express Eq.(\ref{eqn:5}) as:
\begin{align}\label{eqn:9}
\begin{split}
  \textit{H}_3=\sum_{\textit{\textbf{q}},\xi}\sum_{\textit{\textbf{q}}',\xi'}\sum_{\textit{\textbf{q}}'',\xi''}&C_+(\textit{\textbf{q}}\xi,\textit{\textbf{q}}'\xi', \textit{\textbf{q}}''\xi'')
 Q \binom{\textit{\textbf{q}}}{\xi}Q \binom{\textit{\textbf{q}}'}{\xi'}Q \binom{\textit{\textbf{q}}''}{\xi''}^*. 
\end{split}
\end{align}
In Eq.(\ref{eqn:9}), we merge $V_+(\textit{\textbf{q}}j_{\rho},\textit{\textbf{q}}'j'_{\rho'}, \textit{\textbf{q}}''j''_{\rho''})$ and $\Delta(\textit{\textbf{q}}+ \textit{\textbf{q}}'- \textit{\textbf{q}}'')$  into one coefficient $C_+(\textit{\textbf{q}}\xi,\textit{\textbf{q}}'\xi', \textit{\textbf{q}}''\xi'')$. Here we use $\xi$ instead of $j_{\rho}$ to specify each phonon branch at each $\textit{\textbf{q}}$ for brevity. Similarly, for phonon emission processes (one phonon splits into two phonons):
\begin{align} \label{eqn:10}
\begin{split}
  \textit{H}_3=\sum_{\textit{\textbf{q}},\xi}\sum_{\textit{\textbf{q}}',\xi'}\sum_{\textit{\textbf{q}}'',\xi''}&C_-(\textit{\textbf{q}}\xi,\textit{\textbf{q}}'\xi', \textit{\textbf{q}}''\xi'')
 Q \binom{\textit{\textbf{q}}}{\xi}Q \binom{\textit{\textbf{q}}'}{\xi'}^*Q \binom{\textit{\textbf{q}}''}{\xi''}^*. 
\end{split}
\end{align}
$C_-(\textit{\textbf{q}}j_{\rho},\textit{\textbf{q}}'j'_{\rho'}, \textit{\textbf{q}}''j''_{\rho''})$ is the product of $V_-(\textit{\textbf{q}}j_{\rho},\textit{\textbf{q}}'j'_{\rho'}, \textit{\textbf{q}}''j''_{\rho''})$ and $\Delta(\textit{\textbf{q}}- \textit{\textbf{q}}'- \textit{\textbf{q}}'')$, where
\begin{align}\label{addeqn:11}
\begin{split}
   V_-(\textit{\textbf{q}}j_{\rho},\textit{\textbf{q}}'j'_{\rho'}, \textit{\textbf{q}}''j''_{\rho''})&= 
 \sum_{\kappa,\alpha}   \sum_{l,\kappa',\beta}\sum_{l',\kappa'',\gamma} \frac{\Phi_{\alpha \beta \gamma}(0\kappa, l\kappa{'}, l{'}\kappa{''})}{\sqrt{M_\kappa M_{\kappa'} M_{\kappa''}}}\\
\times&e_{\alpha}^{\kappa} \binom{\textit{\textbf{q}}}{j_{\rho}}e_{\beta}^{\kappa'} \binom{\textit{\textbf{q}}'}{j'_{\rho'}}^*e_{\gamma}^{\kappa''} \binom{\textit{\textbf{q}}''}{j''_{\rho''}}^* e^{-i \textit{\textbf{q}}' \cdot \textit{\textbf{R}}_{l}-i \textit{\textbf{q}}''\cdot \textit{\textbf{R}}_{l'}}
\end{split}
\end{align}
\begin{align}\label{addeqn:12}
\Delta(\textit{\textbf{q}}- \textit{\textbf{q}}'- \textit{\textbf{q}}'')=\left\{
\begin{array}{rcl}
1       &      & {\textit{\textbf{q}}- \textit{\textbf{q}}'- \textit{\textbf{q}}''=\textit{\textbf{K}}}\\
0     &      & {\textrm{otherwise}}
\end{array} \right.
\end{align}
\section{phonon scattering selection rules in symmorphic groups}
In this paper, we focus mainly on symmorphic space groups, which do not contain symmetry operations with fractional translations, such as screw axes and glide planes. In symmorphic space groups, the space group $G$ can be expressed as the semi-direct product of the point group $G_0$ and its normal subgroup $A$, which for most cases is the translational group $T$ \cite{Lax,BCC,MEB}:
\begin{align} \label{eqn:11}
G=A \wedge G_0.
\end{align}
Given a phonon wave vector $\textit{\textbf{q}}$, all the group elements $\{R\}$ in $G_0$ that satisfy the following relation:
\begin{align}
R\textit{\textbf{q}}= \textit{\textbf{q}}+\textit{\textbf{K}},
\end{align}
will form a subgroup of the point group, which is called the group of the wavevector $G_0(\textit{\textbf{q}})$. Here $\textit{\textbf{K}}$ is any reciprocal lattice vector.

Next we will apply the selection-rule theory for space groups to the phonon absorption process. It can be shown\cite{latticedynamics} that $Q \binom{\textit{\textbf{q}}}{\xi}$ transforms in the same way as the phonon eigenvector $\textit{\textbf{e}}\binom{\textit{\textbf{q}}}{\xi}$.
Suppose $\textit{\textbf{e}}\binom{\textit{\textbf{q}}}{\xi}$ transforms according to a certain irreducible representation $D^{(\textit{\textbf{q}})(\xi)}_G$ of the group of the wavevector $G_0(\textit{\textbf{q}})$, then the triple product $Q \binom{\textit{\textbf{q}}}{\xi}Q \binom{\textit{\textbf{q}}'}{\xi'}Q \binom{\textit{\textbf{q}}''}{\xi''}^* $ must transform according to the direct product of the representations:
\begin{align}\label{eqn:13}
D^{(\textit{\textbf{q}})(\xi)}_G \otimes D^{(\textit{\textbf{q}}')(\xi')}_G \otimes
D^{(\textit{\textbf{q}}'')(\xi'')^*}_G.
\end{align}

A necessary and sufficient condition for a particular triple product of the complex normal coordinates to appear in the expansion of $H_3$ [in another word, the coefficient $C_+$ is nonzero for a particular triple product in Eq.(\ref{eqn:5})] is that the corresponding character reduction coefficient after the Clebsch–Gordan series expansion is nonvanishing \cite{latticedynamics,MEB}, i.e. in the expansion:
\begin{align}\label{eqn:14}
D^{(\textit{\textbf{q}})(\xi)}_G \otimes D^{(\textit{\textbf{q}}')(\xi')}_G \otimes
D^{(\textit{\textbf{q}}'')(\xi'')}_G= \sum_{n}\langle \textit{\textbf{q}} \xi \otimes \textit{\textbf{q}}'\xi'
\otimes \textit{\textbf{q}}''\xi''^*|n \rangle \Gamma^{n},
\end{align}
$\langle \textit{\textbf{q}} \xi \otimes \textit{\textbf{q}}'\xi'
\otimes \textit{\textbf{q}}''\xi''^*|1 \rangle \neq 0$. Here $n$ labels different irreducible representations and $\Gamma^1$ is the identity representation. For symmorphic space groups, $\langle \textit{\textbf{q}} \xi \otimes \textit{\textbf{q}}'\xi'
\otimes \textit{\textbf{q}}''\xi''^*|1 \rangle$ can be calculated by the following formula\cite{MEB}:
\begin{align}\label{eqn:15}
\langle \textit{\textbf{q}} \xi \otimes \textit{\textbf{q}}'\xi'
\otimes \textit{\textbf{q}}''\xi''^*|1 \rangle &=\frac{1}{g'_0}\sum_{ \{R \}}\chi^{\textit{\textbf{q}} \xi}(\{R\})
\chi^{\textit{\textbf{q}}' \xi'}(\{R\})
\chi^{ \textit{\textbf{q}}'' \xi''}(\{R\})^*\Delta(\textit{\textbf{q}}+ \textit{\textbf{q}}'- \textit{\textbf{q}}''),
\end{align}
where $\chi^{\textit{\textbf{q}} \xi}$ is the character of the irreducible representation $D^{(\textit{\textbf{q}})(\xi)}_G$ and $R$ goes through all the common elements in $G_0(\textit{\textbf{q}}),G_0(\textit{\textbf{q}}')$ and $ G_0(\textit{\textbf{q}}'')$. These symmetry operations form a new subgroup, which we denote as  $G_0(\textit{\textbf{q}},\textit{\textbf{q}}',\textit{\textbf{q}}'')$ and ${g'_0}$ is the number of elements in this group. Eq. (\ref{eqn:15}) can be used to determine whether a particular phonon scattering process is forbidden by the crystal symmetry once the groups of the wavevector $G_0(\textit{\textbf{q}})$, $G_0(\textit{\textbf{q}}')$ and $G_0(\textit{\textbf{q}}'')$ for the three participating phonons are specified. We can write down the condition for a forbidden phonon absorption process as follows:
\begin{align}\label{eqn:16}
\langle \textit{\textbf{q}} \xi \otimes \textit{\textbf{q}}'\xi'
\otimes \textit{\textbf{q}}''\xi''^*|1 \rangle = 0. \quad
\end{align}
Following the same procedure, we can get the similar selection rule for phonon emission processes:
\begin{align}\label{eqn:17}
\langle \textit{\textbf{q}} \xi \otimes \textit{\textbf{q}}'\xi'^*
\otimes \textit{\textbf{q}}''\xi''^*|1 \rangle = 0. \quad
\end{align}
These equations show that once we have determined $G_0(\textit{\textbf{q}},\textit{\textbf{q}}',\textit{\textbf{q}}'')$ and specified the phonon branches, the expansion coefficients $\langle \textit{\textbf{q}} \xi \otimes \textit{\textbf{q}}'\xi'
\otimes \textit{\textbf{q}}''\xi''^*|1 \rangle$ or  $\langle \textit{\textbf{q}} \xi \otimes \textit{\textbf{q}}'\xi'^*
\otimes \textit{\textbf{q}}''\xi''^*|1 \rangle$ can be readily calculated with the help of group character tables using Eq.(\ref{eqn:15}).  

 
\section{application to 2d system: graphene}
In this section, we demonstrate the use of the space-group selection-rule theory elaborated in the previous section by deriving the phonon-phonon scattering selection rules in graphene. Graphene's space group is $P6/mmm$, and the corresponding point group is $D_{6h}$, whose properties are most conveniently studied in terms of the combination of the simpler $C_{6v}$ subgroup that includes all the in-plane symmetries and the out-of-plane mirror symmetry $\sigma_h$. Given graphene's 2D nature, the subgroup $G_0(\textit{\textbf{q}},\textit{\textbf{q}}',\textit{\textbf{q}}'')$ always contains $\sigma_h$, and may have additional in-plane symmetries from $C_{6v}$.
 In the following examples, we will only apply the selection rules for phonon absorption processes since the procedure is the same for phonon emission processes.
\subsection{Selection Rules Imposed by the Mirror Reflection $\sigma_h$}
Since all \textit{\textbf{q}} in the momentum space are confined in a 2D plane, there are at least two elements in $G_0(\textit{\textbf{q}},\textit{\textbf{q}}',\textit{\textbf{q}}'')$:
\begin{align} \label{eqn:18}
 G_0(\textit{\textbf{q}},\textit{\textbf{q}}',\textit{\textbf{q}}'')=\{E,\sigma_h \},  
\end{align}
where $E$ is the identify operation. In this case, $G_0(\textit{\textbf{q}},\textit{\textbf{q}}',\textit{\textbf{q}}'')$ is the point group $C_{1h}$, which means any phonon eigenvector $\textit{\textbf{e}}\binom{\textit{\textbf{q}}}{\xi}$ must transform according to one of its irreducible representations listed in Table 1: $\Delta_1$ or $\Delta_2$. To be concrete, the eigenvectors of in-plane phonon modes (TA, TO and LA, LO modes) remain the same under the mirror reflection $\sigma_h$ and thus belong to the representation $\Delta_1$, while the eigenvectors of out-of-plane phonon modes (the flexural ZA and ZO modes) change sign under $\sigma_h$ and thus belong to the representation $\Delta_2$.
\begin{table}[H]
\large
\centering
\caption{$C_{1h}$ group character table}
\begin{tabular}[t]{l|c|c}
\hline 
&E&$\sigma_{h}$\\
\hline
$\Delta_1$&1&1\\
$\Delta_2$&1&-1\\
\hline
\end{tabular}
\end{table}%

Now consider the scattering process involving two in-plane phonons and one out-of-plane phonon assuming the energy and momentum conservation conditions are satisfied. Then the eigenvector $\textit{\textbf{e}}\binom{\textit{\textbf{q}}}{\xi}$  transforms as $\Delta_1$, $\textit{\textbf{e}}\binom{\textit{\textbf{q}}'}{\xi'}$  transforms as $\Delta_1$ and $\textit{\textbf{e}}\binom{\textit{\textbf{q}}''}{\xi''}$ transform as $\Delta_2$. Applying Eq. (\ref{eqn:15}), we have: 
\begin{align} \label{eqn:19}
\begin{split}
&\langle \textit{\textbf{q}} \xi \otimes \textit{\textbf{q}}'\xi'
\otimes \textit{\textbf{q}}''\xi''^*|1 \rangle\\
=&\frac{1}{g'_0}\sum_{\{R\}}\chi^{\textit{\textbf{q}} \xi}(\{R\})
\chi^{\textit{\textbf{q}}' \xi'}(\{R\})
\chi^{ \textit{\textbf{q}}'' \xi''}(\{R\})^*\\
=&\frac{1}{2}[\chi_{\Delta_1}(E)\chi_{\Delta_1}(E)\chi_{\Delta_2}(E)+\chi_{\Delta_1}(\sigma_{h})\chi_{\Delta_1}(\sigma_{h})\chi_{\Delta_2}(\sigma_{h})]\\
=&\frac{1}{2}[1 \times 1 \times 1+1 \times 1 \times (-1)]\\
=&0,
\end{split}
\end{align}
so this transition is forbidden. This calculation indicates that two in-plane modes cannot scatter into an out-of-plane mode. Similarly, we can repeat the calculation for other possible phonon combinations, which concludes that any three-phonon scattering channels involving an odd number of out-of-plane phonon modes are forbidden in 2D materials with the out-of-plane mirror reflection symmetry $\sigma_h$. These selection rules were first analyzed by Lindsay et al.\cite{lindsay2010flexural} by an explicit symmetry analysis of the scattering matrix elements in graphene, and are responsible for the reduced scattering of the flexural phonons in graphene. Here we justify them through our group theory approach. Similarly, the group theory approach can also reproduce the phonon angular momentum selection rules in 1D chain systems with a screw axis as discussed in previous work\cite{pandey2018symmetry}. We detail the derivation process in Appendix A.

\subsection{Selection Rules Imposed by In-plane Symmetries} 

Besides the mirror reflection $\sigma_h$, the in-plane symmetries of graphene, including the six-fold rotational axis and in-plane mirror reflections, impose additional selection rules on phonon-phonon scattering. For phonon modes with a generic wavevector $\mathbf{q}$ in the momentum space (not located at any high-symmetry lines or points), the associated group of wavevector $G_0(\mathbf{q})$ is trivial and only contains the identity operation and $\sigma_h$. Therefore, for any scattering process involving such a generic phonon mode, the in-plane crystal symmetry will not impose additional selection rules. In other words, the in-plane crystal symmetries will only lead to additional selection rules on scattering processes involving phonon modes located at high-symmetry lines and points, where the associated groups of wavevector contain more symmetry elements and possess nontrivial representations.     

\begin{figure}[h!] \label{fig:1}
    \centering
    \includegraphics[scale=0.4]{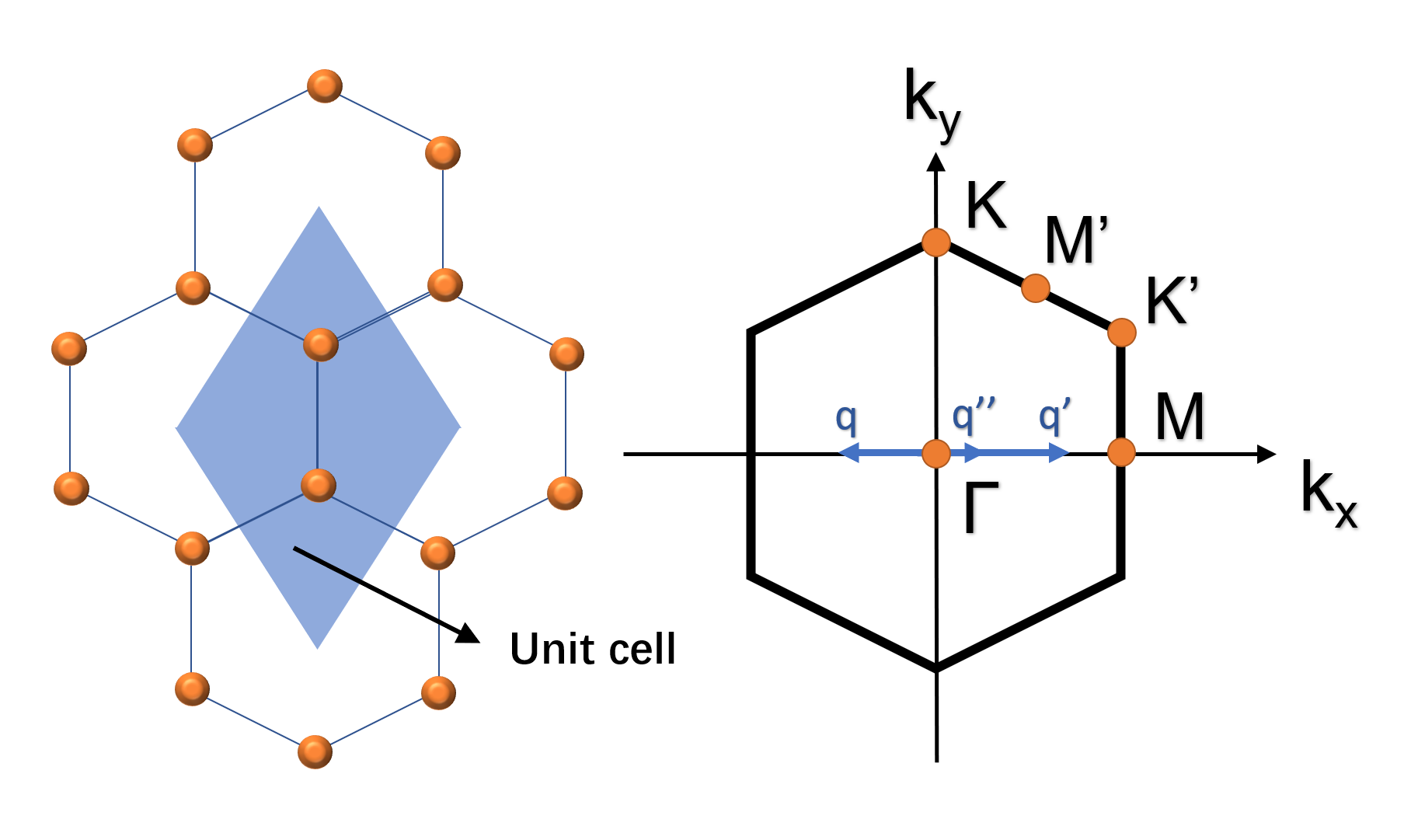}
    \caption{The crystal structure and the Brillouin zone of grpahene. Three phonon modes with wavevectors along the $\Gamma$-M direction in the Brillouin zone of graphene are labeled ($\mathbf{q},\mathbf{q}',\mathbf{q}''$).}
\end{figure}

As one example, consider three phonon modes with wavevectors along the high symmetry path $\Gamma$ to M (as shown in Fig.~1, excluding $\Gamma$ and M points), then $G_0(\textit{\textbf{q}},\textit{\textbf{q}}',\textit{\textbf{q}}'')$ contains four elements:
\begin{align}
 G_0(\textit{\textbf{q}},\textit{\textbf{q}}',\textit{\textbf{q}}'')=\{E,C_2(x),\sigma_{xz},\sigma_h \},  
\end{align}
where $C_2(x)$ is the $\pi$ rotation around the $x$-axis and $\sigma_{xz}$ is the mirror reflection by the $x$-$z$ plane. In this case, $G_0(\textit{\textbf{q}},\textit{\textbf{q}}',\textit{\textbf{q}}'')$ has the same representations as the point group $C_{2v}$,  which means the eigenvector $\textit{\textbf{e}}\binom{\textit{\textbf{q}}}{\xi}$ of any phonon mode along this path must transform according to one of its irreducible representations listed in Table II ($\Delta_1$ to  $\Delta_4$). Since all irreducible representations in this case are one-dimensional, the symmetry properties of the phonon eigenvectors can be easily analyzed by examining their sign change under a particular symmetry operation. The typical eigenvectors of different phonon branches along this path and their respective irreducible representations are shown in Fig. 2. 

\begin{table}[ht]
\large
\centering
\caption{$C_{2v}$ character table}
\begin{tabular}[t]{l|c|c|c|c}
\hline
&E&$C_2(x)$&$\sigma_{xz}$&$\sigma_{h}$\\
\hline
$\Delta_1$&1&1&1&1\\
$\Delta_2$&1&1&-1&-1\\
$\Delta_3$&1&-1&1&-1\\
$\Delta_4$&1&-1&-1&1\\
\hline
\end{tabular}
\end{table}%

Consider a scattering channel involving three phonons along the $\Gamma$-M path, whose eigenvectors belong to the irreducible representations $\Delta_A$, $\Delta_B$ and $\Delta_C$, respectively, and we denote this scattering channel as $\Delta_A \odot \Delta_B \odot \Delta_C$. With the help of the character table (Table II), we can enumerate the forbidden scattering channels by applying Eq.(\ref{eqn:15}). Since no phonon modes belong to the representation $\Delta_2$ along the $\Gamma$-M path in graphene, the following scattering channels for phonons along the $\Gamma$-M path are forbidden by the in-plane crystal symmetries in graphene:  
\begin{figure}[h!] \label{fig:2}
    \centering
    \includegraphics[scale=0.4]{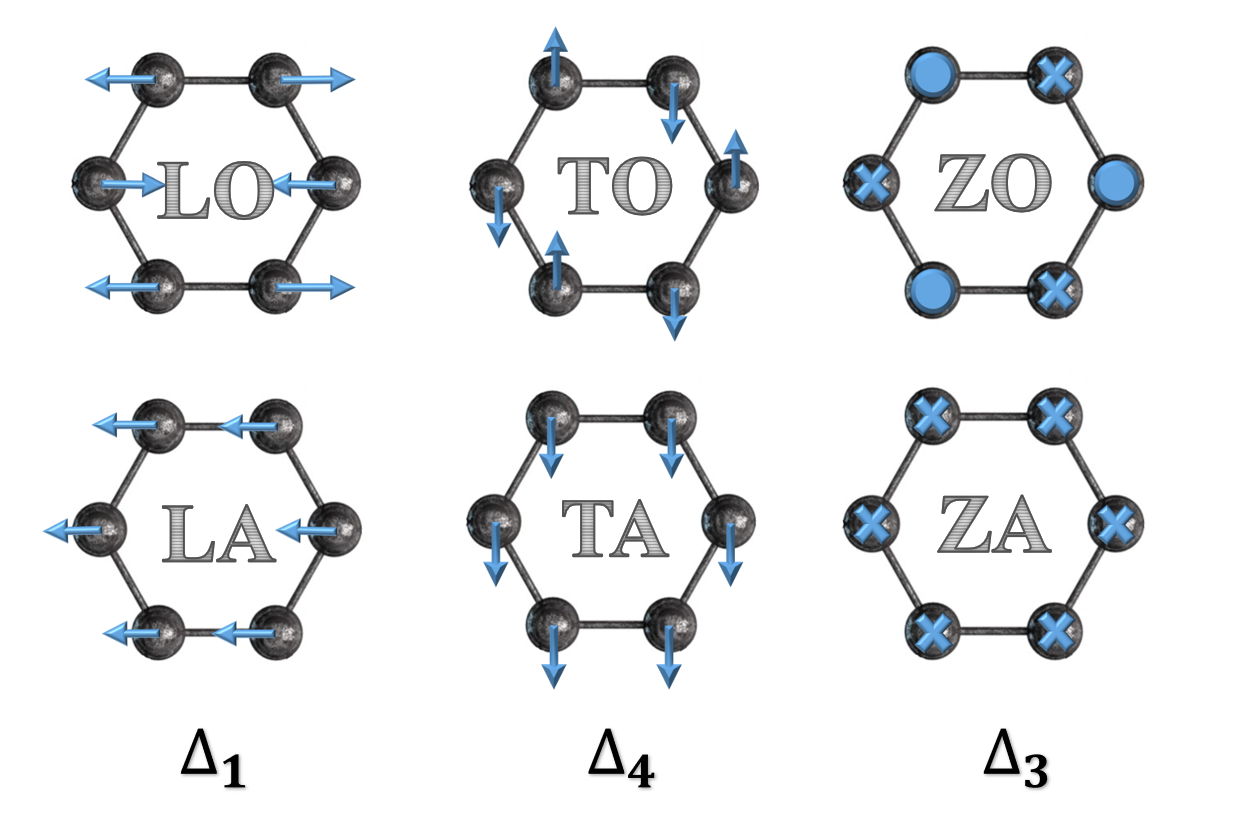}
    \caption{Typical atomic vibrational patterns (eigenvectors) for phonon modes along the $\Gamma$-M path and their corresponding irreducible representation as listed in Table II. The dots and crosses represent directions into and out from the paper plane.}
\end{figure}
\begin{align}
\begin{split}
\Delta_1 \odot \Delta_1 \odot \Delta_3 \quad
\Delta_1 \odot \Delta_1 \odot \Delta_4\\
\Delta_3 \odot \Delta_3 \odot \Delta_3 \quad
\Delta_3 \odot \Delta_3 \odot \Delta_4\\
\Delta_4 \odot \Delta_4 \odot \Delta_3 \quad
\Delta_4 \odot \Delta_4 \odot \Delta_4\\
\Delta_1 \odot \Delta_3 \odot \Delta_4 \quad \quad \quad.
\end{split}
\end{align}
For example, since the channel $\Delta_1 \odot \Delta_1 \odot \Delta_4$ is forbidden, so two longitudinal modes cannot scatter with a transverse mode along the $\Gamma$-M direction, even if the energy and momentum conservation conditions are satisfied. These additional scattering selection rules are imposed strictly by the crystal symmetry, indicating more restrictions on phonon scattering channels and thus higher lattice thermal conductivity for materials with higher crystal symmetries. To gain more physical insights, we explicitly analyze the symmetry properties of the force constants that are associated with the forbidden channels in Appendix B. 

Similarly, we can examine the scattering processes involving phonon modes alont other high symmetry paths. For example, if all three phonon wavevectors are along the high symmetry path $\Gamma$ to K (Fig. 3a), then $G_0(\textit{\textbf{q}},\textit{\textbf{q}}',\textit{\textbf{q}}'')$ has four elements:
\begin{align}
 G_0(\textit{\textbf{q}},\textit{\textbf{q}}',\textit{\textbf{q}}'')=\{E,C_2(y),\sigma_{yz},\sigma_h \}.  
\end{align}
We note here that there are other possible combinations of phonon modes with the same $G_0(\textit{\textbf{q}},\textit{\textbf{q}}',\textit{\textbf{q}}'')$ structure, as long as each mode's group of wavevector contains the same four elements. One example is shown in Fig. 3b. In this case, $G_0(\textit{\textbf{q}},\textit{\textbf{q}}',\textit{\textbf{q}}'')$ has the same representations as the point group $C_{2v}$,  whose character table is given in Table III with irreducible representations $\Delta_1'$ to  $\Delta_4'$. For these phonon modes shown in Fig. 3, the symmetry properties of their eigenvectors and the corresponding irreducible representation that they belong to are listed in Fig. 4. Different from the $\Gamma$-M path, all four representations appear along $\Gamma$-K. Thus, the following forbidden processes can be derived using Eq.(\ref{eqn:15}): 
\begin{figure}[H] \label{fig:3}
    \centering
    \includegraphics[scale=0.4]{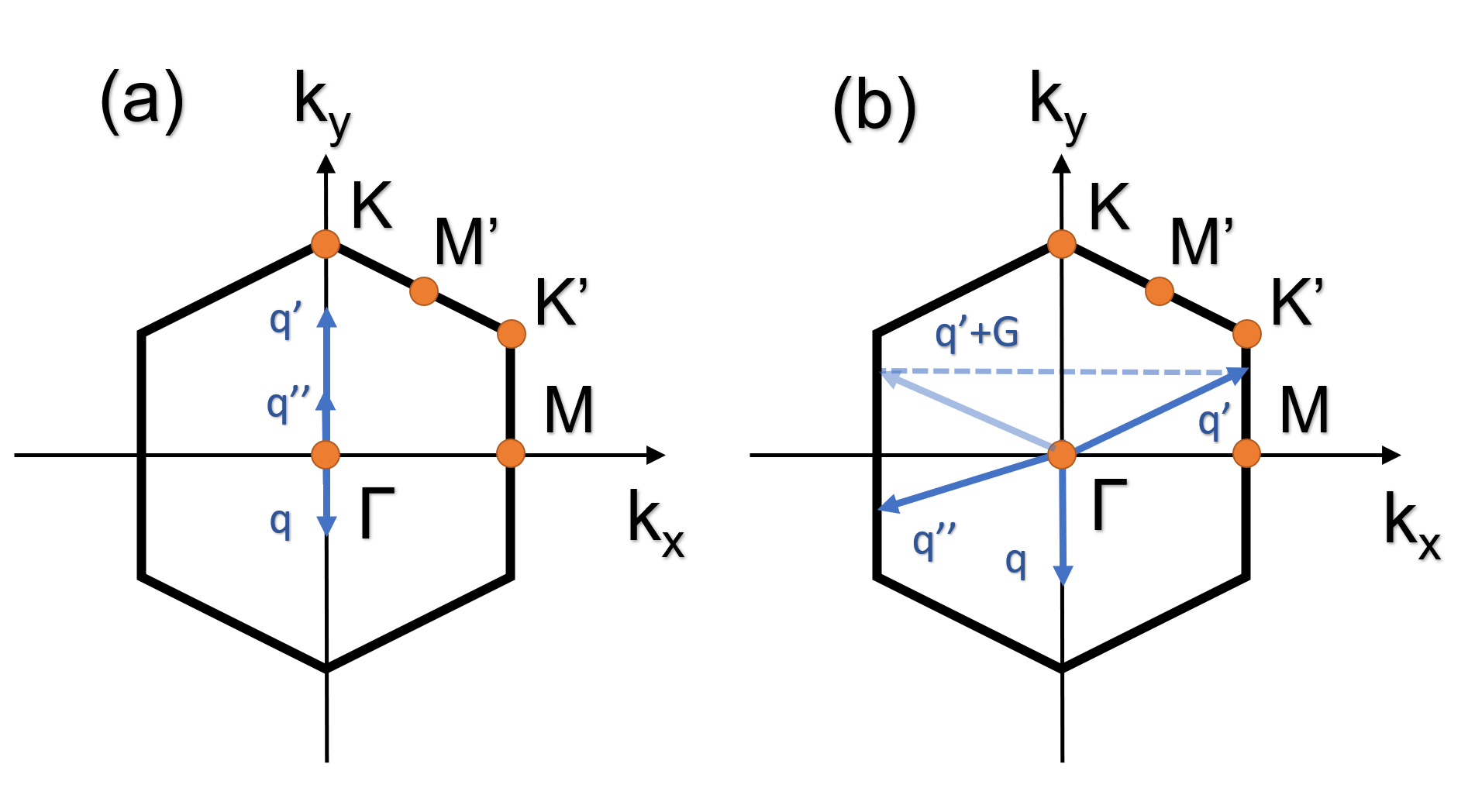}
    \caption{(a) Three phonon modes with wavevectors along the $\Gamma$-K path, whose representations are listed in Table III. (b) One example of three other phonons possessing the same symmetry properties as those along the $\Gamma$-K path.}
\end{figure}

\begin{table}[H]
\large
\centering
\caption{$C_{2v}$ character table}
\begin{tabular}[t]{l|c|c|c|c}
\hline
&E&$C_2(y)$&$\sigma_{yz}$&$\sigma_{h}$\\
\hline
$\Delta_1'$&1&1&1&1\\
$\Delta_2'$&1&1&-1&-1\\
$\Delta_3'$&1&-1&1&-1\\
$\Delta_4'$&1&-1&-1&1\\
\hline
\end{tabular}
\end{table}%

\begin{figure}[h!]\label{fig:4}
    \centering
    \includegraphics[scale=0.3]{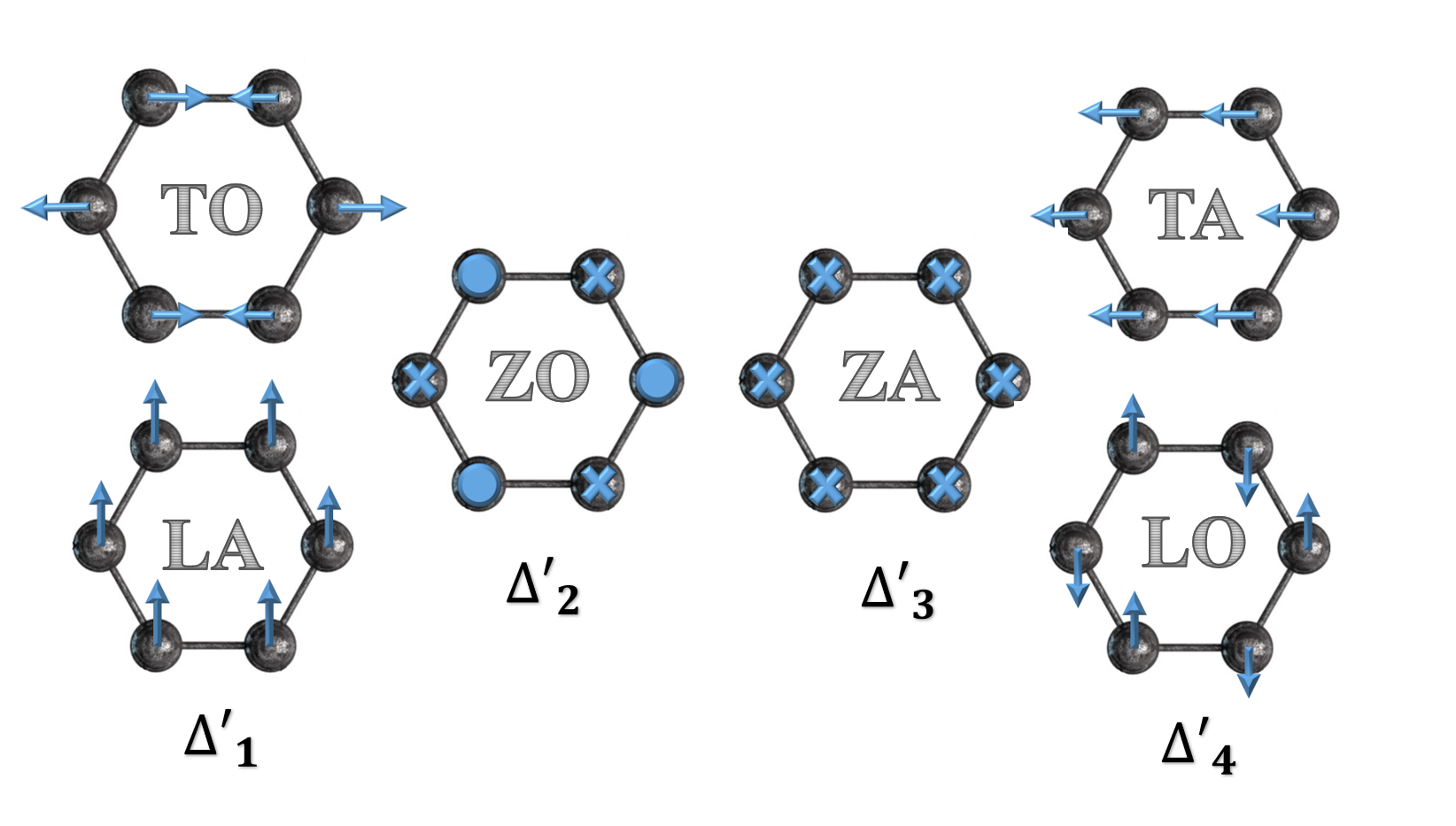}
    \caption{Typical atomic vibrational patterns (eigenvectors) for phonon modes along the $\Gamma$-K path and their corresponding irreducible representation as listed in Table III. The dots and crosses represent directions into and out from the paper plane.}
\end{figure}

\begin{align}
\begin{split}
\Delta_1' \odot \Delta_1' \odot \Delta_2' \quad
\Delta_1' \odot \Delta_1' \odot \Delta_3'\\
\Delta_1' \odot \Delta_1' \odot \Delta_4' \quad
\Delta_2' \odot \Delta_2' \odot \Delta_2'\\
\Delta_2' \odot \Delta_2' \odot \Delta_3' \quad
\Delta_2' \odot \Delta_2' \odot \Delta_4'\\
\Delta_3' \odot \Delta_3' \odot \Delta_2' \quad
\Delta_3' \odot \Delta_3' \odot \Delta_3'\\
\Delta_3' \odot \Delta_3' \odot \Delta_4' \quad
\Delta_4' \odot \Delta_4' \odot \Delta_2'\\
\Delta_4' \odot \Delta_4' \odot \Delta_3' \quad
\Delta_4' \odot \Delta_4' \odot \Delta_4'\\
\Delta_1' \odot \Delta_2' \odot \Delta_3' \quad
\Delta_1' \odot \Delta_2' \odot \Delta_4'\\
\Delta_1' \odot \Delta_3' \odot \Delta_4'. \quad \quad \quad
\end{split}
\end{align}
From this list of forbidden scattering channels, we can conclude, for example, TA+TA $\rightarrow$ TA is a forbidden process while LA+LA $\rightarrow$ LA is an allowed process for phonons along the $\Gamma$-K path. We can also observe that the scatterings of phonons along the $\Gamma$-K path are more restricted by the crystal symmetry than those along the $\Gamma$-M path given the larger number of selection rules. 

\section{Impact of Selection Rules on Thermal Transport in Graphene}
In this section, we use the first-principles calculation to verify the phonon scattering selection rules in graphene. From the discussion in the previous section, when the in-plane symmetries exist, the group of wavevectors $G_0(\mathbf{q},\mathbf{q}',\mathbf{q}'')$ could contain many elements, so there will be more selections rules for phonon scattering. On the other hand, once the crystal symmetries are broken in the $x$-$y$ plane, the group of wavevectors $G_0(\mathbf{q},\mathbf{q}',\mathbf{q}'')$ will only contain $E$ and $\sigma_h$, so most of the selection rules will be lifted (except for the ones imposed by $\sigma_h$) as long as energy and momentum conservation conditions are met. 

In order to directly verify this conclusion and evaluate the effect of symmetry breaking on thermal transport in graphene, we simulate a ``skewed'' graphene with first-principles methods\cite{lindsay2016first}, in which the atomic positions are slightly shifted in $x$-$y$ plane to break all the in-plane symmetries. The structures of the original graphene unit cell and the ``skewed''  graphene unit cell are shown in Fig.\ref{fig:graphene2}. We note here that, although the ``skewed'' graphene structure is artificial, it serves as an extreme example to assess the impact of the symmetry-imposed selection rules. Experimentally achievable uniaxial strains can affect some, but not all, of the in-plane symmetries. In practice, a strain gradient can break all in-plane symmetries but it is difficult to theoretically assess its impact on thermal transport due to the broken lattice periodicity. The details for the first-principles computation are given in Appendix C.

\begin{figure}[H]
    \centering
    \includegraphics[scale=0.40]{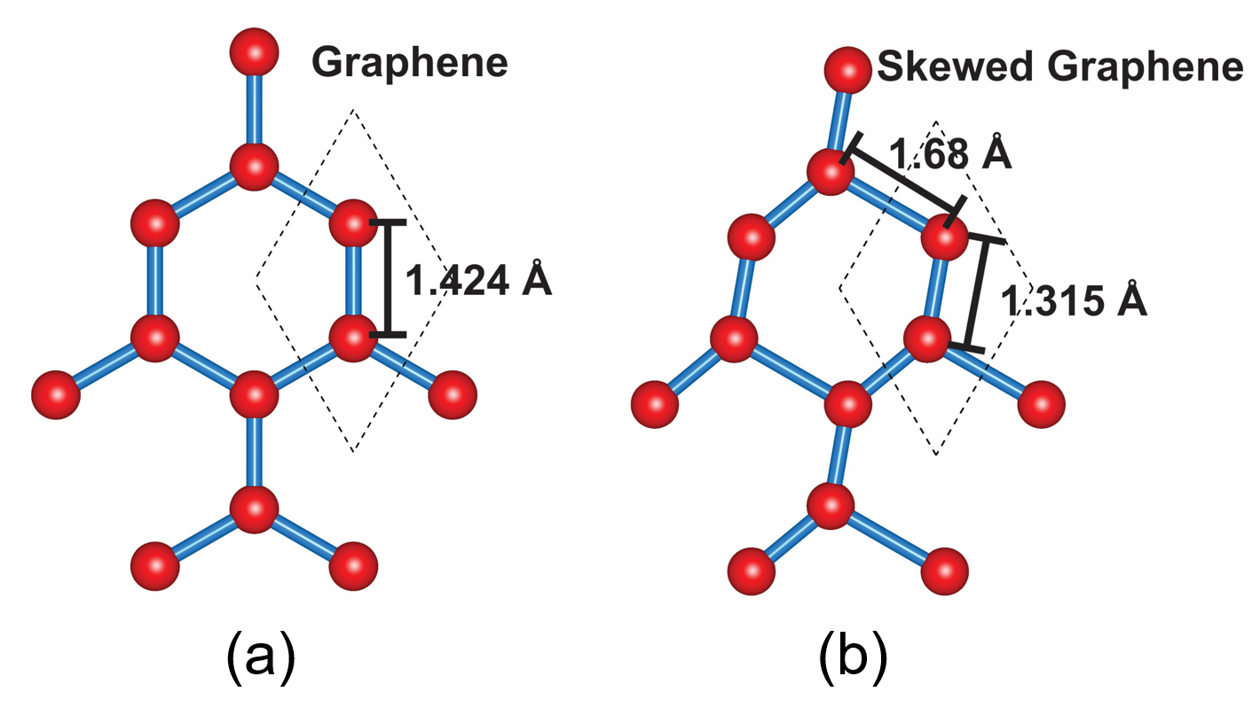}
    \caption{ The crystal structures and unit cells of (a) graphene, and (b) ``skewed'' graphene.}
    \label{fig:graphene2}
\end{figure}
We use first-principles simulation to calculate the mode-specific phonon-phonon scattering rates and decompose them into contributions from different scattering channels. Strictly speaking, the symmetry-imposed selection rules only affect phonon modes located at high-symmetry lines and points in the momentum space. However, it is challenging to only focus on these special phonon modes in the first-principles simulation due to the lack of sufficient data limited by the sampling mesh. So instead, we compute the scattering rates of all phonon modes and examine how the scattering rates of the commonly forbidden scattering channels in graphene (such as TA$\rightarrow$ ZA+ZA and LA+TA $\rightarrow$ LA) change when the in-plane symmetries are broken in the ``skewed'' graphene. In Figs. 6a and 6b, we compare the scattering rates of these two forbidden channels in graphene and ``skewed'' graphene. We find that, for the two forbidden channels in graphene, the phonon-phonon scattering rates increase significantly when the in-plane symmetries are broken in the ``skewed'' graphene. In contrast, for the two allowed channels in graphene (LA $\rightarrow$ ZA+ZA and LA $\rightarrow$ TA+TA), there is no apparent order-of-magnitude change of the phonon scattering rates. We further confirm that the phonon dispersion relations for graphene and ``skewed'' graphene are very similar (Fig. 7a) and these changes in the phonon scattering rates are thus not due to the energy and momentum selection rules. These findings provide verification for the phonon scattering selection rules that we derive using the group theory. Moreover, the broken in-plane crystal symmetry and the lifted scattering selection rules strongly impact the thermal conductivity. As shown in Fig. 7b, the thermal conductivity of the ``skewed'' graphene ($\sim$2100 W/mK at 300 K) is much lower than graphene ($\sim$3000 W/mK at 300 K). Given their similar phonon dispersion relations, the enhanced phonon-phonon scattering due to broken in-plane crystal symmetries leads to a significantly reduction of thermal conductivity ($\sim$30\%) in skewed graphene, illustrating the important influence of crystal symmetry on the lattice thermal conductivity. This result also suggests that the lattice thermal conductivity in high-symmetry materials can potentially be controlled effectively by external conditions that break the crystal symmetry. 

\begin{figure}[H]
    \centering
    \includegraphics[scale=0.6]{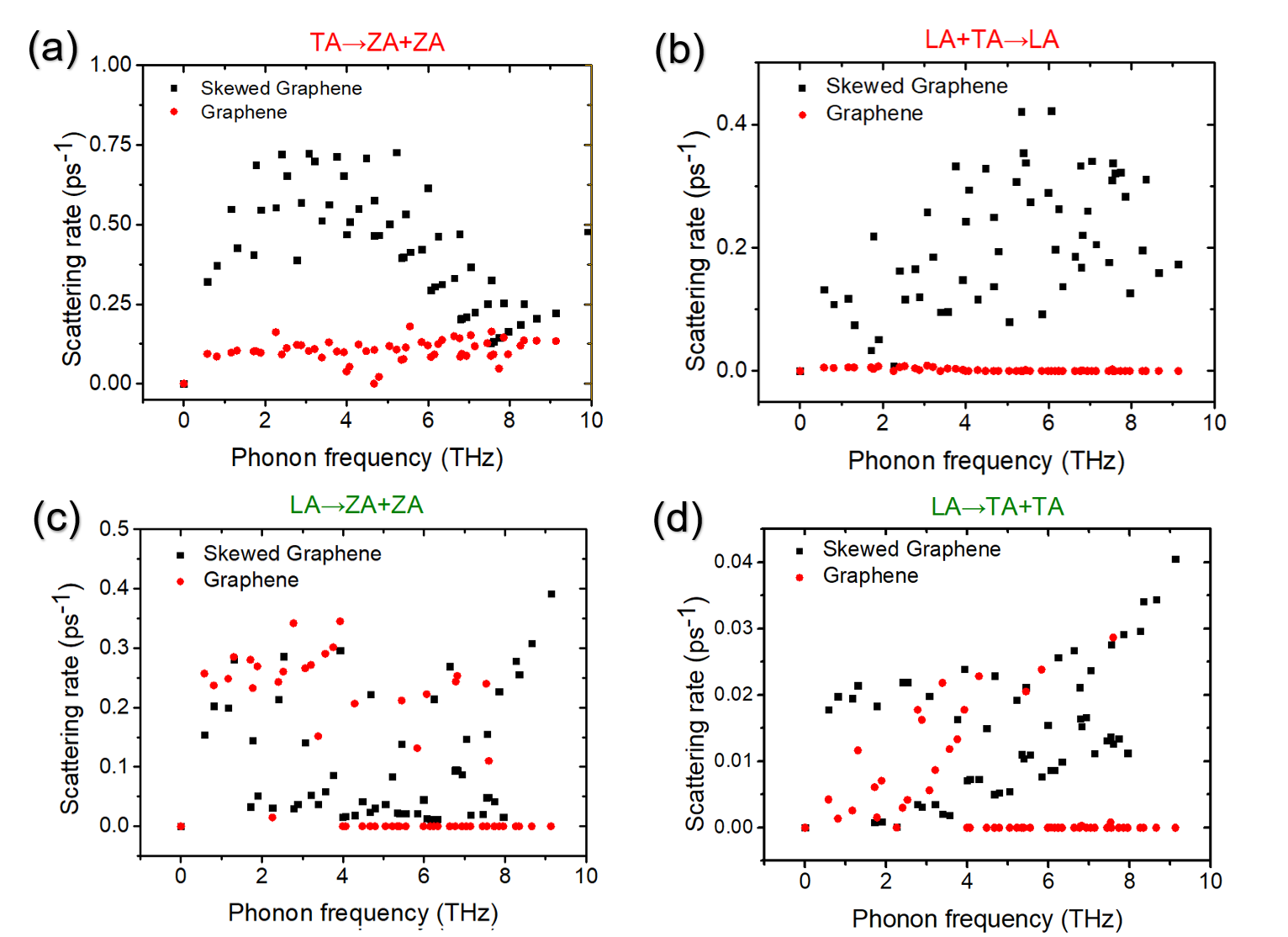}
    \caption{Phonon-phonon scattering rates for both graphene and ``skewed graphene'' decomposed into scattering channels. (a) and (b) show the results for scattering channels that are forbidden by in-plane symmetry in graphene, while (c) and (d) show the results for allowed scattering channels in graphene.}
    \label{fig:graphene1}
\end{figure}

\begin{figure}[H]
    \centering
    \includegraphics[scale=0.5]{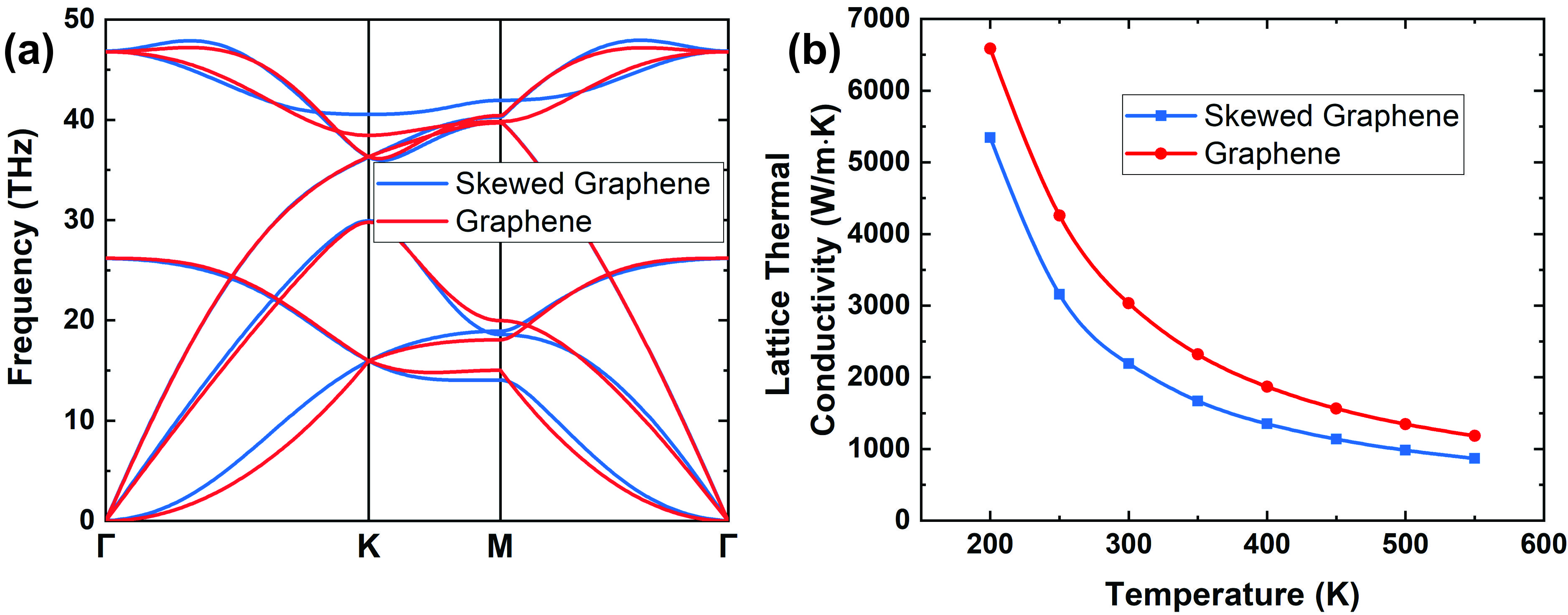}
    \caption{(a) Calculated phonon dispersion relations of graphene and ``skewed'' graphene. (b) Calculated lattice thermal conductivity of graphene and ``skewed'' graphene at different temperatures.}
    \label{fig:graphene2}
\end{figure}
\section{Summary and Discussion}
In summary, we demonstrated that a group-theory-based formalism can be applied to systematically identify anharmonic phonon scattering selection rules imposed by crystal symmetry. We showed that this formalism can reproduce known \textit{ad hoc} selection rules and lead to new ones. We further examined the impact of crystal symmetry breaking on the phonon scattering and thermal transport properties of graphene. Although there has been the qualitative understanding that crystals with higher symmetry tend to possess higher lattice thermal conductivity, our study quantifies the influence of the crystal symmetry. Our recent experimental and computational study of the thermal conductivity of epitaxial gallium arsenide (GaAs) on silicon substrate showed that a small symmetry-breaking in-plane biaxial strain can significantly reduce the lattice thermal conductivity of GaAs\cite{vega2019reduced}. This observation can now be quantitatively understood using the group-theory formalism here, although the derivation is quite tedious so not provided in this paper. 

Furthermore, we note that the group-theory formalism here can be generalized to higher-order scattering processes involving more phonons. For an anharmonic scattering channel involving $N$ phonons, the Clebsch-Gordon coefficient in Eq.(\ref{eqn:15}) can be calculated in a generalized way:
\begin{align}\label{eqn:24}
\begin{split}
\langle N\ \mathrm{phonon} \ \mathrm{process}|1 \rangle &=\large[ \frac{1}{g'_0}\sum_{ \{R \}} \prod_{i=1}^N \chi^{\textit{\textbf{q}}_i \xi_i}(\{R\})^\ddagger\large]
\Delta(\sum_{j=1}^N (-1)^{M_j} \textit{\textbf{q}}_j)\\
\chi^{\textit{\textbf{q}}_i \xi_i}(\{R\})^\ddagger&=\left\{
\begin{array}{rcl}
\large
\chi^{\textit{\textbf{q}}_i \xi_i}(\{R\})       &      &  \textrm{if} \ M_i=0\\
\chi^{\textit{\textbf{q}}_i \xi_i}(\{R\})^*     &      &  \textrm{if} \ M_i=1
\end{array} \right.\\
\Delta(\sum_{j=1}^N (-1)^{M_j} \textit{\textbf{q}}_j)&=\left\{
\begin{array}{rcl}
\large
1       &      &  \textrm{if} \ \sum_{j=1}^N (-1)^M_j \textit{\textbf{q}}_j=\textit{\textbf{K}}\\
0     &      & {\textrm{otherwise}}
\end{array} \right.\\
\textrm{\textit{R} goes through all }& \textrm{elements in the group}  \  G_0(\textit{\textbf{q}}_1, \textit{\textbf{q}}_2, \dots,\textit{\textbf{q}}_N).
\end{split}
\end{align}
While the application of this formula is cumbersome for general crystal symmetries, useful conclusions can be drawn for simple symmetries. For example, it can be shown from Eq.(\ref{eqn:24}) that the forbidden scattering channels involving an odd number of flexural phonons imposed by $\sigma_h$ in graphene still applies to higher-order phonon scattering processes, which was also discussed by Lindsay et al.\cite{lindsay2010flexural}.

Compared to materials with symmorphic space groups, it is more difficult to obtain straightforward selection rules for materials with nonsymmorphic space groups due to the complexity of their symmetry considerations. This suggests that further research can be done to generalize the current formalism to lattices with nonsymmorphic space groups. 

On a practical level, our result suggests that symmetry-breaking strain should be minimized for applications where efficient heat dissipation is desirable\cite{vega2019reduced}. On the other hand, breaking the crystal symmetry in a controlled way can serve as a means to tune the thermal conductivity for solid-state thermal switching applications. For this purpose, a strain gradient can be more effective than uniaxial or biaxial strains.   
\section*{Acknowledgement}
This work is based on research supported by the National Science Foundation under the award number CBET-1846927. R. Y. acknowledges the support of a Chancellor's Fellowship from UCSB. We acknowledge the use of the Center for Scientific Computing supported by the California NanoSystems Institute and the Materials Research Science and Engineering Center (MRSEC) at University of California, Santa Barbara through NSF awards DMR-1720256 and CNS-1725797. 

\appendix
\renewcommand{\thefigure}{A\arabic{figure}}
\setcounter{figure}{0}

\renewcommand{\thetable}{A-\arabic{table}}
\setcounter{table}{0}

\renewcommand{\theequation}{A-\arabic{equation}}
\setcounter{equation}{0}

\section*{Appendix A: Scattering Selection Rules in 1D Chain System with a Screw Axis}
For 1D helical systems with screw axes such as carbon nanotubes and Ba$_3$N, it is known that anharmonic phonon scattering is further subjected to selection rules on the phonon angular momentum\cite{pandey2018symmetry}. Here we analyze this selection rule using the group-theory formalism with Ba$_3$N as an example\cite{pandey2018symmetry}, which contains 1D-chain structures with a two-fold screw axis. The space group of the Ba$_3$N chain can be treated as a symmorphic group because the helical symmetry group containing the screw axis operation is its normal subgroup. In this case, since all phonon wave vectors are along the $z$ direction, the group $G_0(\textit{\textbf{q}},\textit{\textbf{q}}',\textit{\textbf{q}}'')$ is the direct product of the helical symmetry group $H$ and $C_{3v}$ group. The character table is given in Table A-I. Here $S$ is the screw axis operation ($\pi$ rotation around the chain axis plus a half-period translation along the chain direction), $C_3(z)$ is a three-fold rotation around the chain axis ($z$ axis), and $\sigma_v$ is a mirror reflection operation. Typical phonon eigenvectors for the acoustic branches and the corresponding representations are given in Fig. A1. 
\begin{figure}[H]
    \centering
    \includegraphics[scale=0.40]{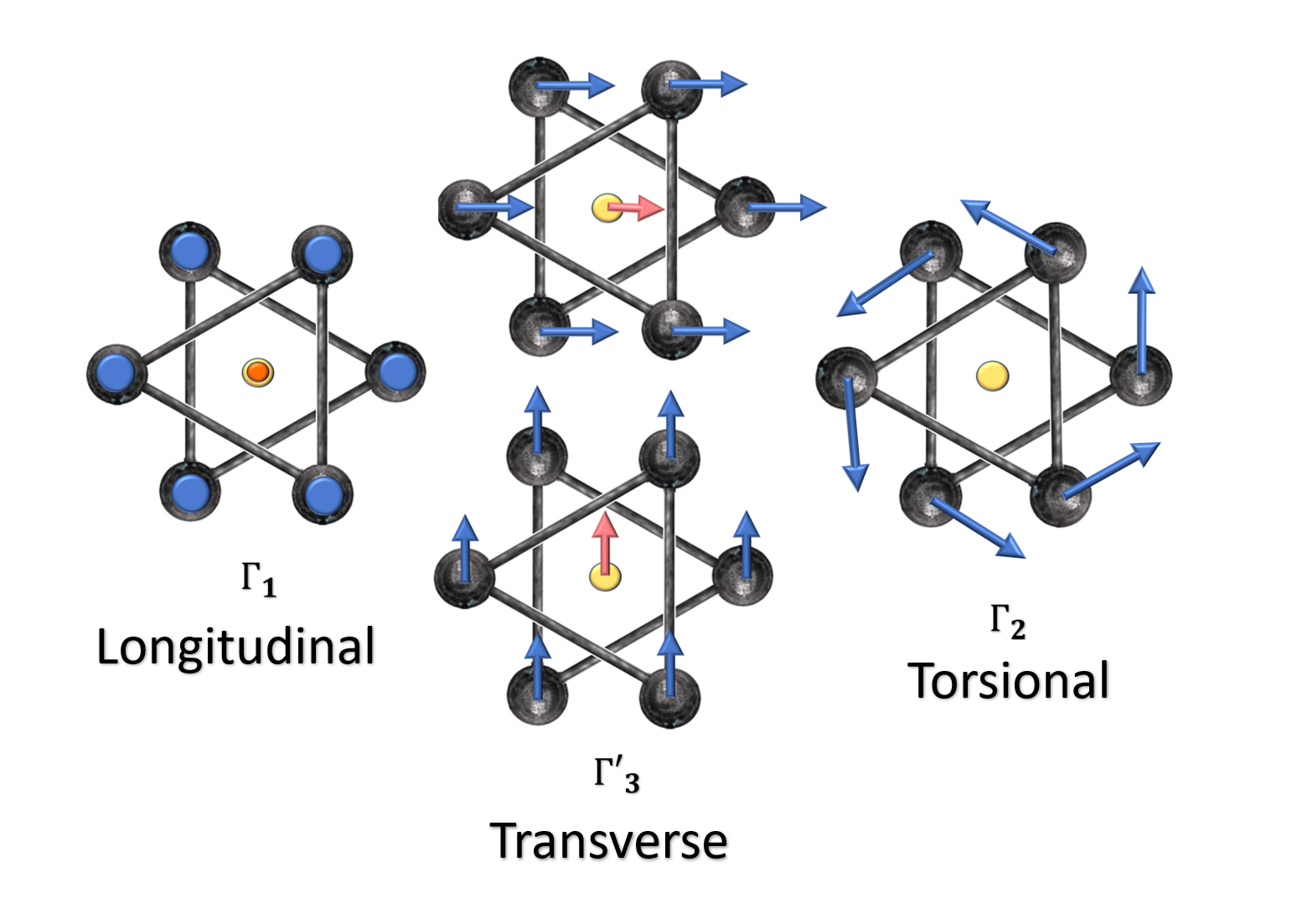}
    \caption{Typical atomic vibrational patterns (eigenvectors) for phonon modes along the 1D Ba$_3$N chains and their corresponding irreducible representation as listed in Table A-I. }
    \label{fig:1dacoustic}
\end{figure}
\begin{table}[H]
\caption{$C_{3v} \otimes H$ character table}
\large
\centering
\begin{tabular}[t]{l|c|c|c|c|c|c}
\hline
&$E$&$2C_3(z)$&$3\sigma_{v}$&$S$&$2SC_3(z)$&$3S\sigma_{v}$\\
\hline
$\Gamma_1$&1&1&1&1&1&1\\
$\Gamma_1'$&1&1&1&-1&-1&-1\\
$\Gamma_2$&1&1&-1&1&1&-1\\
$\Gamma_2'$&1&1&-1&-1&-1&1\\
$\Gamma_3$&2&-1&0&2&-1&0\\
$\Gamma_3'$&2&-1&0&-2&1&0\\
\hline
\end{tabular}
\end{table}

Due to the space limitation, we will not write down all forbidden processes here. Still, we can retrieve some general findings:
the representations without the prime can be seen to have chiralily 0 and those with the prime has chirality 1, depending on how they transform under $S$. Then the selection rules originated from the phonon chirality (or the conservation of phonon angular momentum) discussed by Pandey et al.\cite{pandey2018symmetry} can be easily checked. For instance, using Eq.(\ref{eqn:15}), we can show that the scattering channel
\begin{align}\label{eqn:gai31}
\begin{split}
(\Gamma_1 \ \textrm{or} \  \Gamma_2) + (\Gamma_1 \ \textrm{or} \  \Gamma_2) \rightarrow \Gamma_3'\\
\end{split}
\end{align}
is forbidden, which means two longitudinal or torsional phonon modes cannot scatter into a transverse phonon mode. This is consistent with the phonon angular momentum rule since the left-hand side has a total angular momentum of 0 while the right-hand side has a total angular momentum of 1.
Also, we can identify the following forbidden transitions:
\begin{align}\label{eqn:gai31}
\begin{split}
\Gamma_1 \odot \Gamma_1 \odot \Gamma_2\\
\Gamma_2 \odot \Gamma_2 \odot \Gamma_2\\
\end{split}
\end{align}
which means, if we only consider scatterings among acoustic phonons, two torsional modes cannot be scattered into one longitudinal mode,
and three torsional modes cannot scatter with each other. These results are all consistent with the phonon angular momentum selection rule.

\section*{Appendix B: Symmetry Analysis of the Force Constants}
Here we provide an explicit symmetry analysis of the relevant force constants to understand the selection rules discussed in the main text. For this purpose, we rewrite Eq.(\ref{eqn:6}) in the following form:
\begin{align} \label{eqn:27}
\begin{split}
   &V_+(\textit{\textbf{q}}\Delta_A,\textit{\textbf{q}}'\Delta_B, \textit{\textbf{q}}''\Delta_C)\\
   =&
 \sum_{0\kappa,\alpha}   \sum_{l,\kappa',\beta}\sum_{l',\kappa'',\gamma} \frac{\Phi_{\alpha \beta \gamma}(0\kappa, l\kappa{'}, l{'}\kappa{''})}{\sqrt{M_\kappa M_{\kappa'} M_{\kappa''}}}\\
\times&e_{\alpha}^{\kappa} \binom{\textit{\textbf{q}}}{\Delta_A}e_{\beta}^{\kappa'} \binom{\textit{\textbf{q}}'}{\Delta_B}e_{\gamma}^{\kappa''} \binom{\textit{\textbf{q}}''}{\Delta_C}^* e^{i \textit{\textbf{q}}' \cdot \textit{\textbf{R}}_{l}-i \textit{\textbf{q}}''\cdot \textit{\textbf{R}}_{l'}},
\end{split}
\end{align}
where $\Delta_A$, $\Delta_B$ and $\Delta_C$ are the corresponding representations of $\xi$, $\xi'$ and $\xi''$ in the character table of $G_0(\textit{\textbf{q}},\textit{\textbf{q}}',\textit{\textbf{q}}'')$. Consider the particular forbidden channel for phonons along the $\Gamma$-M path in graphene: $\Delta_1 \odot \Delta_1 \odot \Delta_4$ (two longitudinal phonons scatter with one transverse phonon). One term in the summation in Eq.(\ref{eqn:27}) has the form:
\begin{align}\label{eqn:28}
\begin{split}
V_1=\frac{\Phi_{\alpha \beta \gamma}([0]2, [1]1,[3]2)}{\sqrt{M_c M_c M_c}}
e_{\alpha}^{2} \binom{\textit{\textbf{q}}}{\Delta_1}e_{\beta}^{1} \binom{\textit{\textbf{q}}'}{\Delta_1}e_{\gamma}^{2} \binom{\textit{\textbf{q}}''}{\Delta_4}^* e^{i \textit{\textbf{q}}' \cdot \textit{\textbf{R}}_{[1]}-i \textit{\textbf{q}}''\cdot \textit{\textbf{R}}_{[3]}}.
\end{split}
\end{align}
Here the third-order force constants involve three atoms, whose unit cell indices and atomic indices inside each unit cell are labeled in Fig. A2. For each term $V_1$, there is always a corresponding term $V_2$ in the summation with the following form:
\begin{figure}[h] \label{fig:A2}
    \centering
    \includegraphics[scale=0.55]{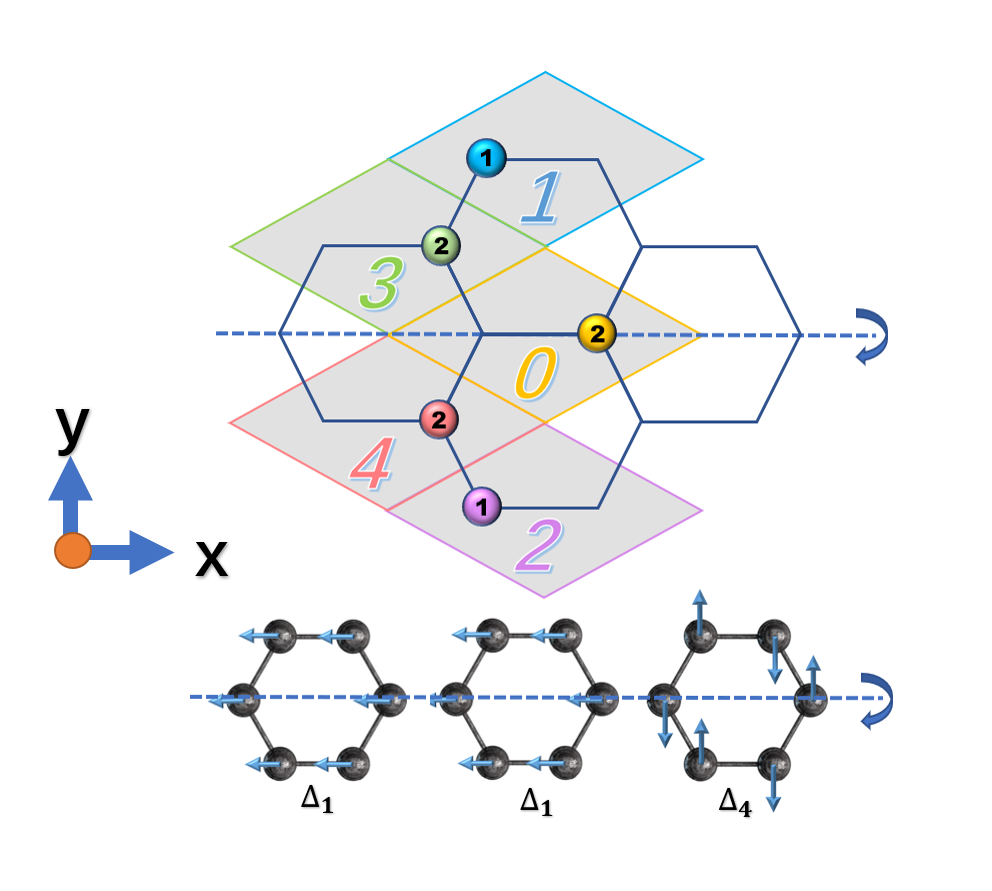}
    \caption{An illustration of the graphene crystal structure, where the three atoms involved in the force constants used in Eq.(\ref{eqn:28}) and Eq.(\ref{eqn:A5}) are labeled. The vibrational modes of the three participating phonons are also illustrated.}
     \label{justi1a}
\end{figure}

\begin{align}\label{eqn:A5}
\begin{split}
V_2=\frac{\Phi_{\alpha \beta \gamma}([0]2, [2]1,[4]2)}{\sqrt{M_c M_c M_c}}
&e_{\alpha}^{2} \binom{\textit{\textbf{q}}}{\Delta_1}e_{\beta}^{1} \binom{\textit{\textbf{q}}'}{\Delta_1}e_{\gamma}^{2} \binom{\textit{\textbf{q}}''}{\Delta_4}^*  e^{i \textit{\textbf{q}}' \cdot \textit{\textbf{R}}_{[2]}-i \textit{\textbf{q}}''\cdot \textit{\textbf{R}}_{[4]}}\\
\end{split}
\end{align}
Given the reflection symmetry about the $x$-axis, it is obvious that:
\begin{align}
e^{i \textit{\textbf{q}}' \cdot \textit{\textbf{R}}_{[1]}-i \textit{\textbf{q}}''\cdot \textit{\textbf{R}}_{[3]}}=
e^{i \textit{\textbf{q}}' \cdot \textit{\textbf{R}}_{[2]}-i \textit{\textbf{q}}''\cdot \textit{\textbf{R}}_{[4]}}.
\end{align}
Further, we know that the force constants are related by crystal symmetries in the following form \cite{lindsay2010flexural,symmlue}:
\begin{align}
\begin{split}
\sum_{\alpha' \beta' \gamma'}\Phi_{\alpha' \beta' \gamma'}([0]2, [1]1, [3]2)
\Omega_{\alpha'\alpha}\Omega_{\beta'\beta}\Omega_{\gamma'\gamma} = \Phi_{\alpha \beta \gamma}([0]2, [2]1, [4]2). 
\end{split}    
\end{align}
Here $\Omega_{\alpha'\alpha}$ are scalar elements of a matrix representing a symmetry operation. If we take the symmetry operation to be the mirror reflection about the $x$-axis (the direction of phonon wavevectors), the above equation gives:
 $\Phi_{xxy}([0]2, [1]1, [3]2)
= -\Phi_{xxy}([0]2, [2]1, [4]2)$. Since $\Delta_1$ modes only have polarizations along the $x$-direction and $\Delta_4$ modes only along the $y$-direction, $\Phi_{xxy}$ are the only force constants contributing to $V_1$ and $V_2$. Therefore, we have:
\begin{align}
V_1+V_2=0.  
\end{align}
The above analysis shows that for any give term $V_1$ in the summation in Eq. (\ref{eqn:27}), it is always possible to find another term $V_2$ that the sum of these two terms is zero given the reflection symmetry by the $x$-axis. Thus, the total sum in Eq.(\ref{eqn:27}), namely $V_+=0$, must be zero, confirming that this particular scattering channel is forbidden. Other selection rules derived in the main text can be understood in a similar manner.

\section*{Appendix C: Details for the First-Principles Calculation}
We applied the first-principles calculations to obtain the phonon properties of the normal and the ``skewed'' graphene. The Vienna Ab-initio Simulation Package (VASP)\cite{dft1,dft2} based on density functional theory (DFT) were adopt for all simulations. The Perdew-Burke-Ernzerhof (PBE) of generalized gradient approximation (GGA) was chosen as the exchange correlation functional\cite{dft3}. We used the projector augmented wave (PAW) potentials\cite{dft4,dft5} to describe the core (1s$^2$) electrons, with the 2s$^2$ and 2p$^2$ electrons of carbon considered as valence electrons\cite{dft6}. The kinetic energy cutoff of wave functions was set at 500 eV, \cite{dft6} and Monkhorst-pack k-mesh $20\times20\times1$ including the $\Gamma$ point was used to sample the Brillouin zone for both cases. Vacuum layers with $10 \AA$ thickness were used to hinder the self-interactions between atomic layers arising from the periodic boundary condition. In the calculation of phonon dispersions, $5 \times 5 \times 1$ supercells were constructed for both structures. To obtain the phonon dispersions, we calculated the second-order interatomic force constants employing the finite displacement method using PHONOPY package\cite{dft7}. 
For the calculations of the lattice thermal conductivity, the anharmonic third-order force constants were computed using the same supercell and k-mesh. Interactions among atoms up to the fifth nearest neighbor were taken into account. The convergence of the interaction distance was checked. With the second and third order force constants, we solved the phonon Boltzmann transport equation (BTE) by an iterative method using ShengBTE package\cite{dft8}. The q-grid mesh density for converged thermal conductivities in both cases was $100 \times 100 \times 1$.

\bibliography{main}

\end{document}